\documentstyle[12pt,psfig,oldlfont]{article}
\newcommand{\lsim}{\raisebox{-0.13cm}{~\shortstack{$<$ \\[-0.07cm] $\sim$}}~}

\newcommand{\ra}{\rightarrow}
\newcommand{\ee}{e^+e^-}
\newcommand{\s}{\\ \vspace*{-3.5mm} }
\newcommand{\nn}{\noindent}
\newcommand{\non}{\nonumber}
\newcommand{\beq}{\begin{eqnarray}}
\newcommand{\eeq}{\end{eqnarray}}

\newcommand{\MSSM}{\mbox{${\cal MSSM}$}}

\newcommand{\tb}{\tan\beta}

\def\inpb{\mbox{$\hbox{pb}^{-1}$}}

\font\ninerm=cmr9

\begin{document}

\begin{titlepage}


\begin{flushright}
CERN PPE/96-34\\
DESY 95--212\\
IFT--96--05\\
KA--TP--08--96\\
\end{flushright}

\def\thefootnote{\fnsymbol{footnote}}

\vspace{1cm}

\begin{center}

{\large\sc {\bf  SUSY Decays of Higgs Particles}}

\vspace{1cm}

{\sc A.~Djouadi$^{1,2}$\footnote{Supported by Deutsche Forschungsgemeinschaft
DFG (Bonn).}, P. Janot$^3$, J. Kalinowski$^{4}$\footnote{Supported in part by
Committee for Scientific Research and by the EC under contract
CHRX-CT92-0004.} and P.M.~Zerwas$^1$ }

\vspace{.5cm}

{\ninerm
$^1$ Deutsches Elektronen--Synchrotron DESY, D--22603 Hamburg/FRG. \\

$^2$ Inst. Theor. Physik, Univ. Karlsruhe, D--76128 Karlsruhe/FRG. \\

$^3$ PPE Division, CERN, CH--1211, Geneva 23/Switzerland.\\

$^4$ Institute of Theoretical Physics,  Warsaw University, PL--00681
Warsaw/Poland.}

\end{center}

\vspace{.9cm}

\begin{abstract}
\normalsize
\noindent

\nn Among the possible decay modes of Higgs particles into
supersymmetric states, neutralino and chargino decays play a prominent
r\^ole. The experimental opportunities of observing such decay modes at
LEP2 and at future $\ee$ linear colliders are analyzed
within the frame of the Minimal Supersymmetric extension of the
Standard Model. For heavy Higgs
particles, the chargino/neutralino decay modes can be very important,
while only a small window is open for the lightest CP--even Higgs
particle. If charginos/neutralinos are found at LEP2, such decay modes
can be searched for in a small area of the parameter space, and
invisible decays may reduce the exclusion limits of the lightest CP-even
Higgs particle slightly; if charginos/neutralinos are not found at LEP2
in direct searches, the Higgs search will not be affected by the SUSY
particle sector.

\end{abstract}

\end{titlepage}

\normalsize

\def\thefootnote{\arabic{footnote}}
\setcounter{footnote}{0}
\setcounter{page}{2}

\nn {\bf 1.} New light will be shed on the Higgs sector of
supersymmetric theories in the forthcoming LEP2 experiments~\cite{R1}.
In fact, if the parameter $\tb$ in the Minimal Supersymmetric Standard
Model (\MSSM) is realized within the range between 1 and 3 as suggested
by unification of $b$--$\tau$ Yukawa couplings~\cite{R1a},
prospects of discovering at least the lightest CP--even neutral Higgs
boson $h$ at LEP2 are very good. At future $\ee$ linear colliders, the
entire  parameter space can be explored. The lightest Higgs boson $h$
can be discovered with certainty, and the Higgs bosons $H,A$ and $H^\pm$
are accessible in major parts of the parameter space. \s

Most theoretical predictions of Higgs decay properties have focussed in
the past on final states built up by the electroweak gauge bosons,
standard quarks/leptons and cascade decays~\cite{R2}. While decays to
squarks and sleptons are expected to be less important and are in
general forbidden kinematically for fairly light Higgs bosons, decays to
neutralinos and charginos
\beq
h, H, A & \ra & \chi_i^0 \chi_j^0 \ \ {\rm and} \ \ \chi_i^+ \chi_j^- \non,
\\ H^\pm & \ra & \chi_i^0 \chi_j^\pm,
\eeq
[$i,j=1$--$4$ for neutral, and $i,j=1,2$ for charged states],  in
particular to the light $\chi_1^0$ states, could play a potentially
important r\^ole~\cite{R3a,R3b}. The final states would have interesting
topologies. If R--parity is conserved and $\chi_1^0$ is the lightest
supersymmetric stable particle (LSP), the $\chi_1^0 \chi_1^0$ final
states are invisible. In the other modes, $\chi_i^0$ and $\chi_i^\pm$
decay cascades~\cite{R3b,R4} would come with large missing energies in
the events. For instance, the decays $h,A \ra \chi_1^0 \chi_2^0$ could
lead to final states $ \chi_1^0 \chi_1^0 f \bar{f}$, with the two LSPs
escaping undetected. \s

These novel decay modes can only be relevant if the standard decay modes
are not overwhelming. Prime interest is therefore restricted to low to
moderate $\tb$ values since $b$ and $\tau$ decays are otherwise
dominant. It turns out {\it a posteriori} that
when  $\chi$ decay channels are open, the branching ratios can be
close to 100\%, opening opportunities  for the
experimental analysis of the fundamental parameters in supersymmetric
theories. \s

In this note, possible $\chi \chi$ decays of Higgs particles are
analyzed in the parameter range of LEP2 and future $\ee$ linear
colliders which
should
eventually run
with energies up to 2 TeV, sweeping the entire
\MSSM\ Higgs spectrum up to masses of order 1 TeV~\cite{R5}. These
 unconventional Higgs decays, in
particular the invisible $\chi_1^0 \chi_1^0$ final states, could change
the experimental analyses for discovering \MSSM\  Higgs particles at the
LHC rather dramatically, making the search of these particles much more
difficult. This problem is much less
severe at $\ee$ colliders so that the \MSSM\ Higgs parameter space can
still be covered completely~\cite{R5b}. It would only be more difficult
to detect the CP--odd $A$ particle for small $\tb$ where it is produced
only in association with the light CP--even particle $h$, leading to
invisible events altogether if both particles decay into LSP final
states. However, since the branching ratios for $\chi_1^0 \chi_1^0$ is
always less than unity [especially when other $\chi\chi$ decays are
kinematically possible], such a problem can  be solved by
increasing the integrated luminosity. \s

In the subsequent analysis,
the fact is taken into account
that neither charginos nor neutralinos
have been found at LEP1.5~\cite{LEP1.5}.

\bigskip

\nn {\bf 2.} Besides the conventional parameters of the \MSSM\ Higgs
sector, the $\chi$ masses and couplings to the Higgs particles are
determined by the Higgs--higgsino mass parameter $\mu$ and the SU(2)
gaugino mass parameter $M_2$. \s

The general chargino mass matrix~\cite{R6},
\begin{eqnarray}
{\cal M}_C = \left[ \begin{array}{cc} M_2 & \sqrt{2}M_W \sin \beta
\\ \sqrt{2}M_W \cos \beta & \mu \end{array} \right],
\end{eqnarray}
is diagonalized by two matrices $U$ and $V$, leading to the
$\chi_{1,2}^\pm$ masses
\begin{eqnarray}
m_{\chi_{1,2}^\pm} = && \frac{1}{\sqrt{2}} \left\{ M_2^2+\mu^2+2M_W^2
\right. \non \\
&& \left. \mp \left[ (M_2^2-\mu^2)^2+4 M_W^2 \cos^2 2\beta+4M_W^2 (M^2_2+\mu^2
+2M_2\mu \sin 2\beta) \right]^{\frac{1}{2}} \right\}^{\frac{1}{2}}.
\end{eqnarray}
When one of the two parameters $\mu$ or $M_2$ is large, one of the
charginos corresponds to a pure gaugino state while the other
corresponds to a pure higgsino state; in these cases the $\chi^\pm$
masses approach  the asymptotic values:
\begin{eqnarray}
|\mu| \gg M_Z \ \ {\rm and} \ \ M_2 \sim M_Z \ &:& m_{\chi_{1}^\pm} \sim M_2
\ , \ m_{\chi_{2}^\pm} \sim |\mu| \non; \\
|\mu| \sim M_Z \  \ {\rm and} \ \ M_2 \gg M_Z \ &:& m_{\chi_{1}^\pm} \sim |\mu|
\ , \ m_{\chi_{2}^\pm} \sim M_2.
\end{eqnarray}

The four-dimensional neutralino mass matrix depends on the same two
mass parameters if the SUGRA relation $M_1=\frac{5}{3} \tan^2 \theta_W$ $
M_2 \simeq \frac{1}{2} M_2$~\cite{R6} is used. In the bino--wino--higgsino
basis $(-i\tilde{B}, -i\tilde{W}_3, \tilde{H}^0_1,$ $\tilde{H}^0_2)$
the matrix has the form
\begin{small}
\begin{eqnarray}
{\cal M}_N = \left[ \begin{array}{cccc}
M_1 & 0 & -M_Z \sin\theta_W \cos\beta & M_Z  \sin\theta_W \sin\beta \\
0   & M_2 & M_Z \cos\theta_W \cos\beta & -M_Z  \cos\theta_W \sin\beta \\
-M_Z \sin\theta_W \cos\beta & M_Z  \cos \theta_W \cos\beta & 0 & -\mu \\
M_Z \sin\theta_W \sin \beta & -M_Z  \cos\theta_W \sin\beta & -\mu & 0
\end{array} \right].
\end{eqnarray}
\end{small}
\nn It can be diagonalized analytically~\cite{R7} by a single matrix $Z$;
as the final results for the masses are rather involved, they should not
be recorded here. Again, for large values of one of the parameters $\mu$
or $M_2$, two neutralinos are pure gaugino states while the two others
are pure higgsino states, and the $\chi^0$ masses are given
in these limits by
\begin{eqnarray}
|\mu| \gg M_Z \  \ {\rm and} \ \ M_2 \sim M_Z \ &:&
m_{\chi_1^0} \sim M_1 \ , \
m_{\chi_2^0} \sim M_2 \ , \ m_{\chi_3^0} \sim m_{\chi_4^0} \sim |\mu|
\non, \\
|\mu| \sim M_Z \  \ {\rm and} \ \ M_2 \gg M_Z \ &:&
m_{\chi_1^0} \sim m_{\chi_2^0} \sim |\mu| \ , \
\ m_{\chi_3^0} \sim M_1 \ , \ m_{\chi_4^0} \sim M_2.
\end{eqnarray}

A typical set of the neutralino/chargino masses [the lightest of which
could
be observed at LEP2] is shown as a function of $\mu$ in
Fig.1a for $M_2=150$~GeV and $\tb=1.6$. The non--observation of chargino and
neutralino production $\ee \to \chi_1^+\chi_1^-$ and $\chi_2^0\chi_1^0$
at LEP1.5~\cite{LEP1.5} roughly translates to lower limits on
$2m_{\chi_1^+}$ and $m_{\chi_2^0}+m_ {\chi_1^0}$ of 135~GeV, so that
the range $-40~{\rm GeV} \lsim \mu \lsim 140$ GeV is ruled out
for $M_2=150$ GeV and $\tb=1.6$. [For $\mu=0$ and also
for moderate $\mu$ values, the lightest neutralino and chargino  states
could have been massless---these  values are of course
already ruled out by the negative search of chargino states at LEP1.]

\bigskip

\nn {\bf 3.} The decay widths of the Higgs bosons $(H_k)= (H,h,A,H^\pm)$
into neutralino and chargino pairs are given by~\cite{R3a}
\begin{eqnarray}
\Gamma (H_k \ra \chi_i \chi_j) = \frac{G_F M_W^2}{2 \sqrt{2} \pi}
\frac{ M_{H_k} \lambda^{1/2} }{1+\delta_{ij}} \hspace*{-3mm}
&& \left[ (F_{ijk}^2 + F_{jik}^2) \left(1- \frac{ m_{\chi_i}^2}{M_{H_k}^2}
        - \frac{ m_{\chi_j}^2}{M_{H_k}^2} \right) \right. \non \\
&& \left. -4\eta_k \epsilon_i \epsilon_j F_{ijk} F_{jik} \frac{ m_{\chi_i}
m_{\chi_j}} {M_{H_k}^2} \right],
\end{eqnarray}
where $\eta_{1,2,4}=+1$, $\eta_3=-1$ and $\delta_{ij}=0$ unless the
final state consists of two identical (Majorana) neutralinos in which
case $\delta_{ii}=1$; $\epsilon_i =\pm 1$ is the sign of the
neutralino mass eigenvalue [$\epsilon_i=1$ for charginos];
$\lambda=(1- m_{\chi_i}^2/M_{H_k}^2 - m_{\chi_j}^2/M_{H_k}^2)^2- 4
m_{\chi_i}^2 m_{\chi_j}^2/M_{H_k}^4$ is the usual two--body phase space
function. \s

In the case of neutral Higgs boson decays, the coefficients $F_{ijk}$ are
related to the elements of the matrices $U,V$ for charginos and
$Z$ for neutralinos,
\begin{eqnarray}
H_k \ra \chi_i^+ \chi_j^- \ &:& F_{ijk}= \frac{1}{\sqrt{2}} \left[
e_k V_{i1}U_{j2} -d_k V_{i2}U_{j1} \right] \non, \\
H_k \ra \chi_i^0 \chi_j^0 \ \ &:& F_{ijk}= \frac{1}{2}
\left( Z_{j2}- \tan\theta_W
Z_{j1} \right) \left(e_k Z_{i3} + d_kZ_{i4} \right) \ + \ i \leftrightarrow j,
\end{eqnarray}
with the coefficients $e_k$ and $d_k$ given by
\begin{eqnarray}
e_1/d_1=\cos\alpha/-\sin \alpha \ , \
e_2/d_2=\sin\alpha/\cos \alpha \ ,  \
e_3/d_3=-\sin\beta/\cos \beta.
\end{eqnarray}
The partial widths of the neutral Higgs particles for decays to the
lightest neutralino or chargino states can be written in a
particularly simple form,
\begin{eqnarray}
\Gamma (H_k \ra \chi_1 \chi_1) = \frac{G_F M_W^2}{2 \sqrt{2} \pi}
M_{H_k} \left[ 1- \frac{ 4m_{\chi_1}^2}{M_{H_k}^2} \right]^p \
\kappa_k^2,
\end{eqnarray}
with $p=3$ for $h,H$ and $p=1$ for $A$, corresponding to P-- and
S--wave final states as required by the CP=$\pm$ quantum numbers of the Higgs
bosons. \s

For the light scalar and the pseudoscalar Higgs decays to the
lightest neutralinos, $h/A\ra \chi_1^0\chi_1^0$,
the coefficients $\kappa_k$  are given by
\beq
\kappa_h &=& \left( Z_{12}- \tan\theta_W Z_{11} \right) \left(\sin
\alpha Z_{13} + \cos\alpha Z_{14} \right) \non, \\
\kappa_A &=& \left( Z_{12}- \tan\theta_W Z_{11} \right) \left(-\sin
\beta Z_{13} + \cos\beta Z_{14} \right). \label{neut}
\eeq
For the decays to the lightest charginos, $h / A\ra \chi_1^+\chi_1^-$,
the coefficients $\kappa_k$  are of the form
\beq
\kappa_h &=&\frac{1}{\sqrt{2}} \left( \sin\alpha V_{11}U_{12} - \cos\alpha
V_{12}U_{11} \right) \non, \\
\kappa_A &=&\frac{1}{\sqrt{2}} \left( -\sin\beta V_{11}U_{12} +
 \cos\beta V_{12} U_{11} \right) .
\eeq
It follows from Eq.(\ref{neut}) that  the Higgs
bosons couple to mixtures of gaugino and higgsino
components of the lightest neutralino, corresponding to
the matrix elements $Z_{11},Z_{12}$ and
$Z_{13},Z_{14}$ respectively. If the $\chi_1^0$ were either a pure
gaugino or a pure higgsino state, the $h\chi_1^0 \chi_1^0$ and $A
\chi_1^0 \chi_1^0$ couplings would vanish. The matrix elements are
displayed for the same set of parameters as for the neutralino/chargino
masses in Fig.1b. It is obvious from the figure that decays to LSPs play
a less prominent r\^ole for $\mu<0$ than for $\mu>0$ since for negative
$\mu$, the $\chi_1^0$ state tends to be either gaugino-- or
higgsino--like. In this case, only in a small $\mu$ window around $
-M_1$ do the couplings reach the level of a few percent, as is shown in
Fig.1c. [For the values $\tb=1.6$ and $M_A=100$ GeV that were
chosen for illustration, the couplings $\kappa_h$ and $\kappa_A$ are
 approximately
equal.]

\bigskip

For the charged Higgs boson, the decay widths into neutralino/chargino
pairs~\cite{R3a} are  given by  Eq.(7), with
\begin{eqnarray}
F_{ij4} &  = & \cos\beta \left[ Z_{j4} V_{i1} + \frac{1}{\sqrt{2}}
\left( Z_{j2} + \tan \theta_W Z_{j1} \right) V_{i2} \right] \non, \\
F_{ji4} & = & \sin \beta \left[ Z_{j3} U_{i1} - \frac{1}{\sqrt{2}}
\left( Z_{j2} + \tan \theta_W Z_{j1} \right) U_{i2} \right].
\end{eqnarray}

\bigskip

\nn {\bf 4.} Examples of branching ratios are shown in Fig.2a
for Higgs boson masses
$M_h$ and $M_A$ typically accessible at LEP2. The partial decay widths
to R--even particles are treated according to the 
current knowledge~\cite{R2}. Whenever the $\chi_1^0\chi_1^0$ decay is
kinematically allowed, the branching ratio is close to 100\% for
positive $\mu$ values. For $\mu<0$ the branching ratio never exceeds the
20\% level. As pointed out previously, the branching ratios become smaller
for increasing $\tb$, except when $h$ reaches its maximal mass value: the
$hbb$ coupling is no longer enhanced in this case. These $M_h$
values are not accessible at LEP2,
but they would be accessible at future colliders. \s

The branching ratios for the sum into all possible neutralino
and chargino states are shown in Fig.2b for
$(H,A,H^\pm)$ particles which can be produced at high--energy $\ee$
colliders. Mixing~\cite{R8} in the Higgs sector has been treated
properly for $\mu \neq 0$, $A_t=\sqrt{6}M_S$ [so-called ``maximal mixing"]
and $A_b=0$, with $M_S=1$ TeV. These branching ratios are always large
except in three cases: $(i$) For $H$ in the mass range between 140 and
200 GeV, especially if $\mu>0$, due to the large value of BR$(H \ra
hh)$; ($ii$) For small $A$ masses and negative $\mu$ values as discussed
above; and ($iii$) For $H^\pm$ just above the $t\bar{b}$ threshold if
not all the decay channels into  heavy $\chi$ states are open. [A
similar pattern holds for the other values of $\mu$ and $M_2$ discussed
above.] \s

Even above the thresholds of decay channels including top quarks, the
branching ratios for the decays into charginos and neutralinos are
sizeable. For very large Higgs boson masses, they reach a common value
of $ \sim 40\%$ for $\tb =1.6$. In fact, as a consequence of the
unitarity of the $U,V$ and $Z$ matrices, the total widths of the three
Higgs boson decays to charginos and neutralinos do not depend on $M_2$,
$\mu$ or $\tb$ in the asymptotic regime $M_{H_k} \gg m_\chi$,
\begin{eqnarray}
\Gamma (H_k \ra \sum_{i,j} \chi_i \chi_j) = \frac{3G_F M_W^2}{4 \sqrt{2} \pi}
M_{H_k} \left( 1+\frac{1}{3} \tan^2 \theta_W \right),
\end{eqnarray}
giving rise to the branching ratio
\begin{eqnarray}
{\rm BR}( H_k \ra \sum_{i,j} \chi_i \chi_j) = \frac{ \left( 1+\frac{1}{3}
\tan^2 \theta_W \right) M_W^2 }{ \left( 1+\frac{1}{3} \tan^2
\theta_W \right) M_W^2 + m_t^2 \cot^2 \beta + m_b^2 \tan^2 \beta },
\end{eqnarray}
Only the leading $t\bar{t}$, $b\bar{b}$ modes for neutral and the
$t\bar{b}$ modes for the charged Higgs bosons need be included in the
total widths. This branching ratio is shown in Fig.2c as a function of
$\tb$. It is always large, even for extreme values of $\tb \sim 1$ or $50$,
where it still is at the 20\% level.

\bigskip

\nn {\bf 5.}
To discuss the impact of invisible decays on the search of Higgs particles
at LEP2, two cases must be distinguished. \s

$(i)$ The region in the $[\mu,M_2]$ parameter space in
which CP-even $h$ Higgs boson decays into the lightest
neutralino $\chi_1^0 \chi_1^0 $ can be observed
at LEP2, is illustrated in Fig.~3a
for the mass $M_h = 90$ GeV.  The same mixing pattern as before
has been considered: $\mu \neq
0$, $A_t=\sqrt{6}M_S$ and $A_b=0$, with $M_S=1$ TeV.
To find  the region which is experimentally accessible by the search
for invisible decays of the Higgs boson, the product $R_{\rm inv}=
{\mathrm BR}(h \to
\chi_1^0 \chi_1^0) \times \sin^2(\beta-\alpha)$ has to exceed
0.45~\cite{R13a} under the LEP2 luminosity conditions.
The small parameter range in which the invisible Higgs decays could
be detected is shown in Fig.~3a together with the corresponding
$[\mu,M_2]$ range which can be investigated by direct chargino and
neutralino searches at LEP2, $e^+e^-
 \ra \chi_i^+ \chi_j^-$ and $\chi_i^0 \chi_j^0$.
For the latter, a minimum total cross-section of 100~fb~\cite{R9},
relevant for a total integrated luminosity of 150~\inpb\ that should
be delivered each year to the four LEP experiments, was required for large
mass splitting between the lightest neutralino and the charginos and
other neutralinos, and the unavoidable loss of efficiency for small mass
differences was parameterized in a realistic manner~\cite{R10}.  \s

It can immediately be seen from this example that the
region in which the light Higgs boson $h$
could be detected in the  $\chi_1^0\chi_1^0$  decay mode
 is small if the LEP1.5 results are taken into account, and it is
embedded completely in the
area covered by direct neutralino and chargino  searches. Since this
holds true for any Higgs boson mass at all
centre-of-mass energies of LEP2, the $[\mu,M_2]$ area cannot be
extended through the Higgs boson search. \s

$(ii)$ The exclusion limits of the lightest CP--even Higgs boson $h$
are affected by invisible decay modes. This is shown at
$\tan\beta=1.6$ for a few examples of the parameters $\mu$ and $M_2$
in Fig.~3b. It is well-known that LEP2 covers the entire $h$ mass range
for $\tan\beta \lsim 2$ if $h$ decays only into SM particles~\cite{R13a}.
However, it is apparent from the examples in
Fig.~3b that for $\mu$ and $M_2$ values in a small strip, adjacent to the
upper right boundary of the crossed range
in Fig.~3a,
the exclusion limit is significantly smaller than the theoretical
upper Higgs mass limit. For these values of $\mu$ and $M_2$, both
$R_{\rm inv}$ and $R_{\rm vis}={\mathrm BR}(h \to b\bar{b}) \times
\sin^2(\beta-\alpha)$ drop below the lower limits,
given in Figs.31 and 32 of Ref.\cite{R13a},
for which Higgs events could be detected in either of the two channels.
Thus, in a small area of the SUSY parameter space Higgs bosons could
escape undetected under the LEP2 running conditions even for low
$\tan \beta$.

\newpage

\begin{figure}[htbp]
\centerline{\psfig{figure=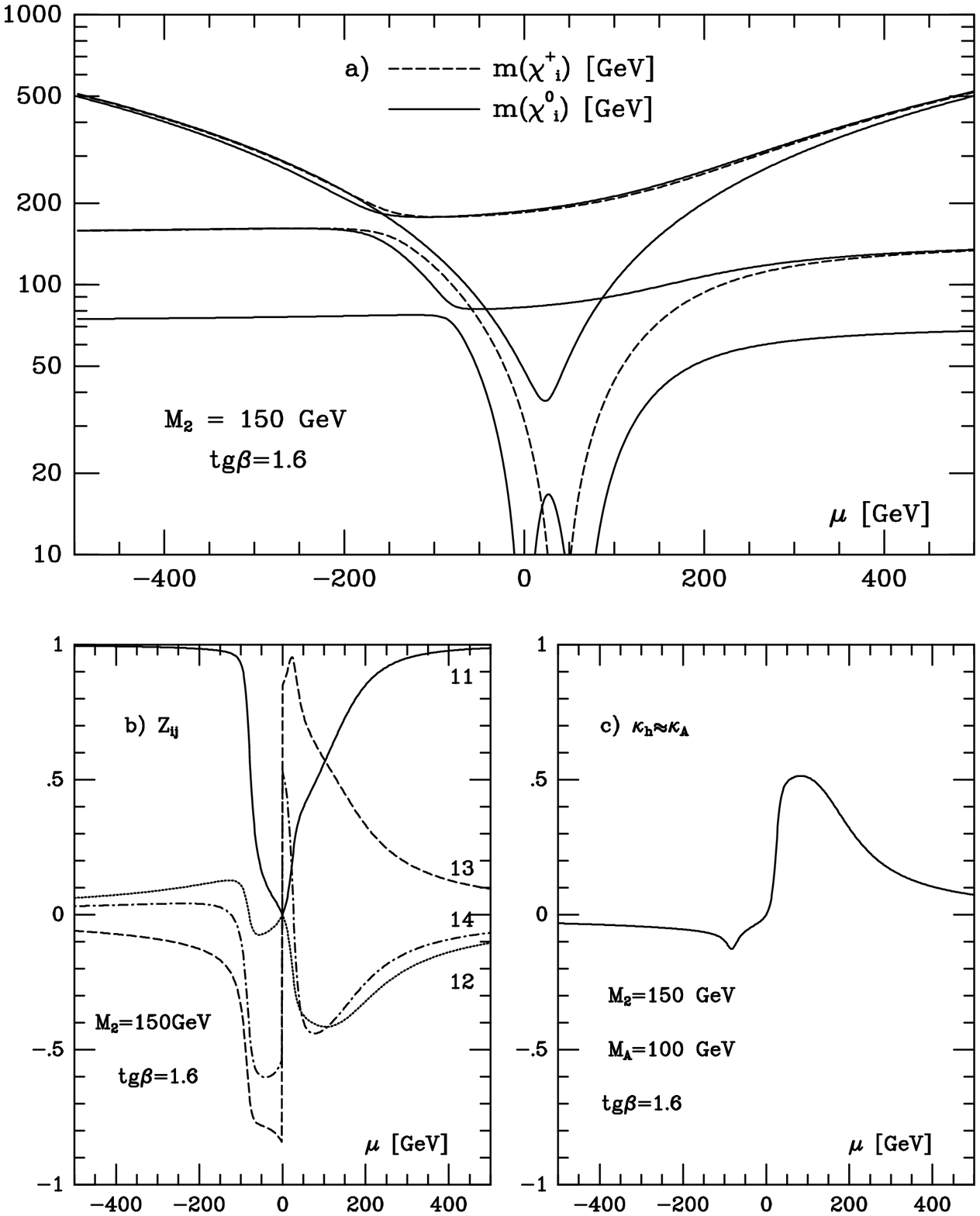,height=20.cm,width=16cm}}
\vspace*{-1.2cm}
\noindent {\bf Fig.~1:} {\it (a) Chargino [dashed lines] and neutralino
[solid lines] masses as a function of $\mu$; (b) the mixing matrix
elements $Z_{1i}$ of the lightest neutralino as a function of $\mu$; (c)
the couplings $\kappa_h$ and $\kappa_A$ of the $h,A$ Higgs bosons to
$\chi_1^0$ as a function of $\mu$. $M_2$ is fixed to 150 GeV and $\tb=1.6$.}
\end{figure}

\newpage

\begin{figure}[htbp]
\centerline{\psfig{figure=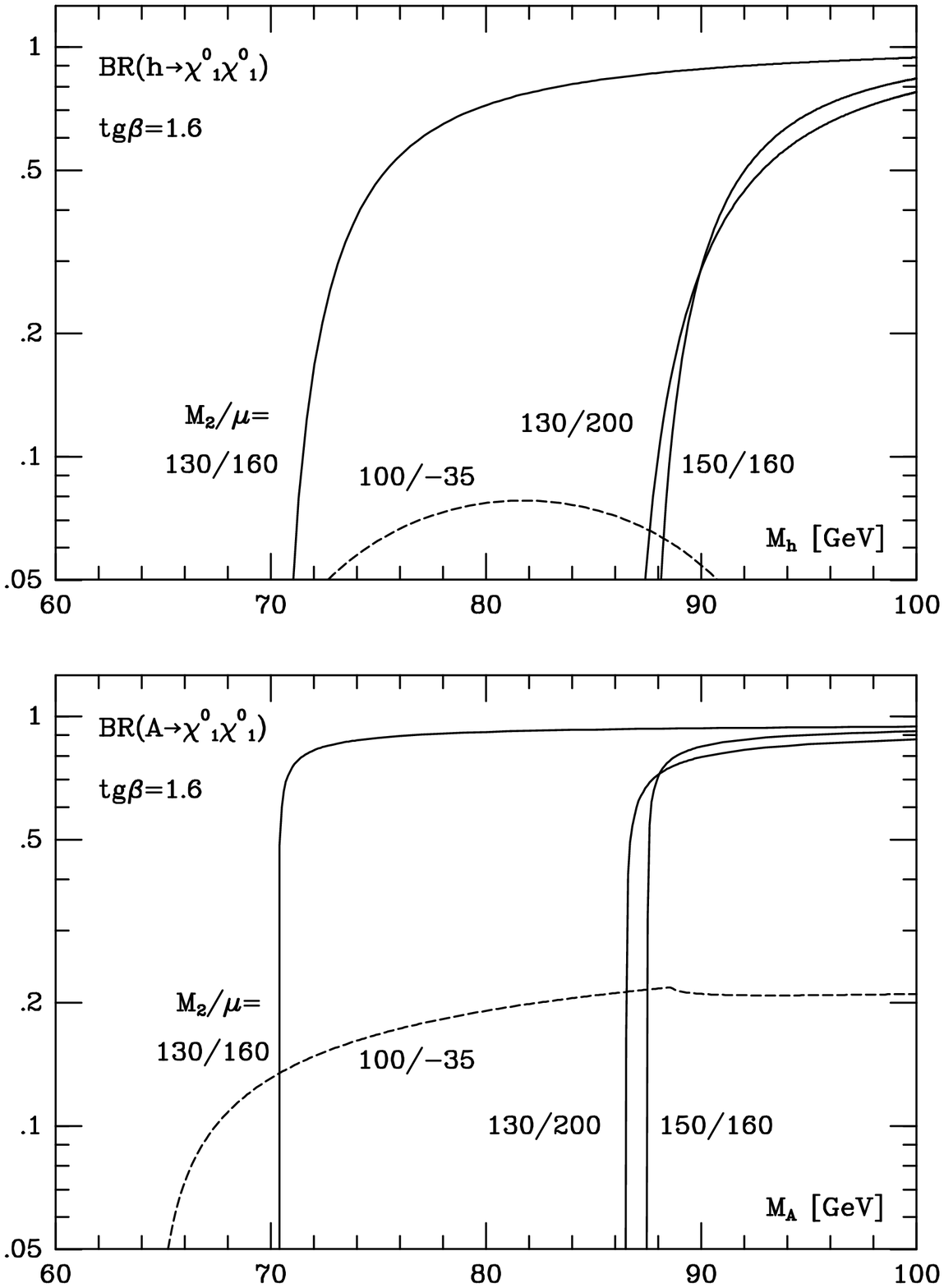,height=20cm,width=16cm}}
\vspace*{-1.2cm}
\noindent {\bf Fig.~2a:} {\it Branching ratios of the decays of the $h$
and $A$
bosons into the lightest neutralino pair $\chi_1^0 \chi_1^0$ as a function of
the Higgs masses for a set of $\mu$ and $M_2$ values; $\tb=1.6$.}
\end{figure}

\newpage

\begin{figure}[htbp]
\centerline{\psfig{figure=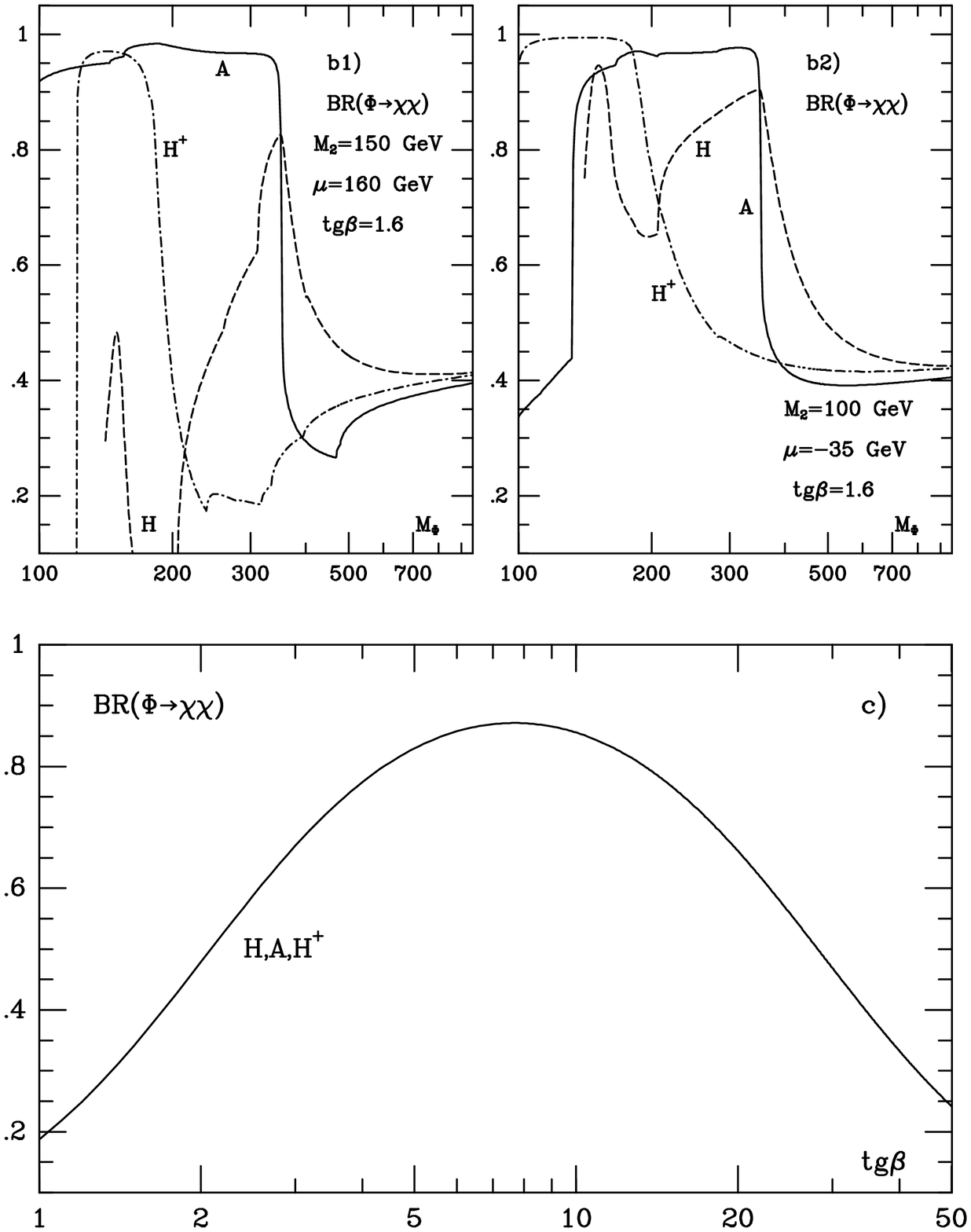,height=20cm,width=16cm}}
\vspace*{-1.2cm}
\noindent {\bf Fig.~2b/c:} {\it Branching ratios of the decays of the heavy
$A$ [solid], $H$ [dashed] and $H^\pm$ [dot--dashed] Higgs bosons into the sum
of neutralino and chargino pairs as a function of the Higgs mass [for a set of
$\mu/M_2$ values and $\tb$ fixed to 1.6] in (b1) and (b2). The inclusive
$\chi \chi$ decay branching ratio as a function of $\tb$ in the
asymptotic region [$M_A \sim M_H \sim M_{H^\pm} = 1$ TeV $\gg m_{\chi}$]
in (c). }
\end{figure}

\newpage

\begin{figure}[htbp]
\centerline{\psfig{figure=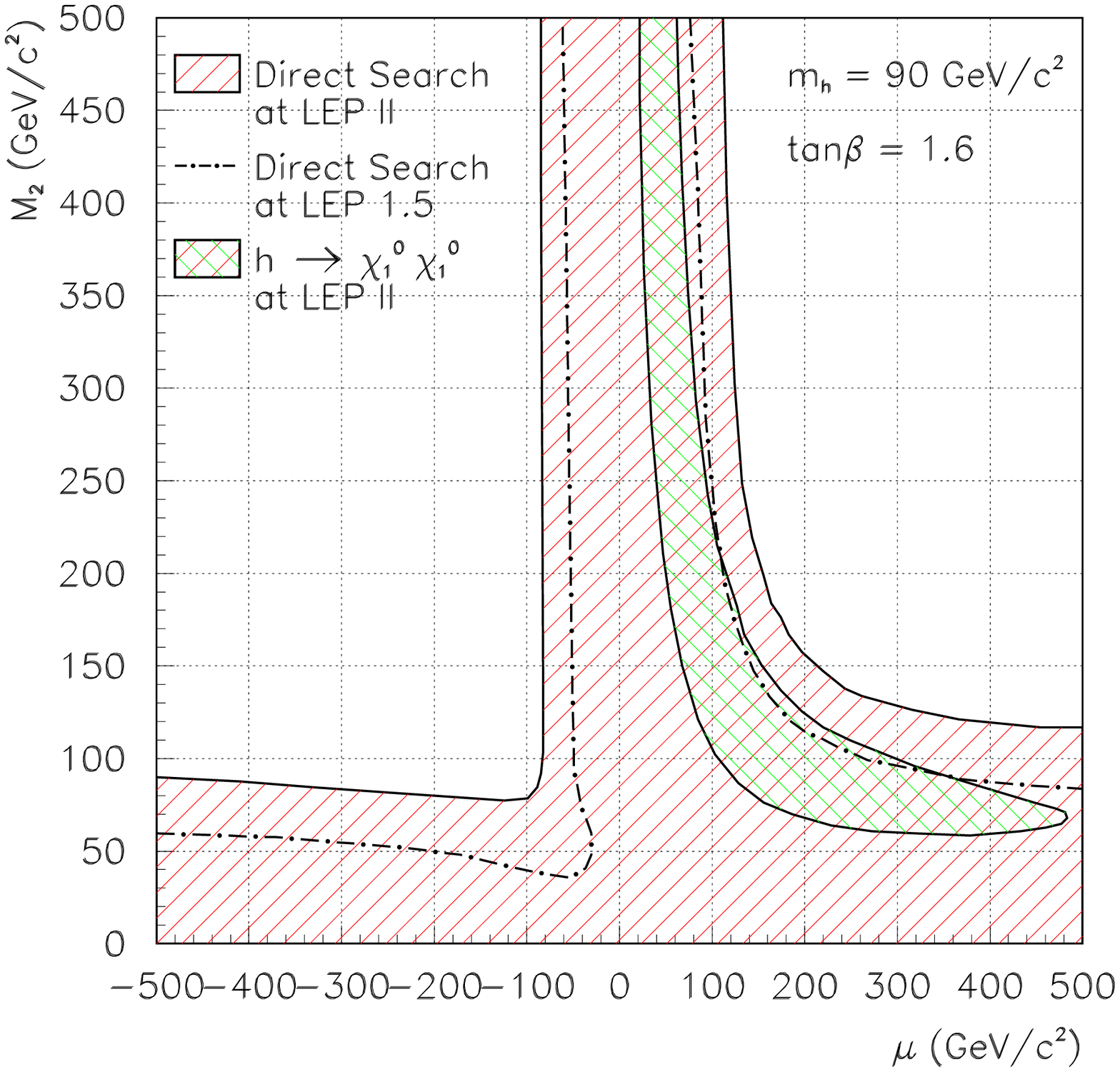,height=16cm,width=16cm}}
\noindent {\bf Fig.~3a:} {\it In the $[\mu, M_2]$ parameter space, ranges
covered at LEP2, with a centre-of-mass of 192~GeV and an integrated luminosity
of 150~\inpb\ delivered to each of the four experiments, by the direct
chargino and neutralino searches, and by the search for invisible Higgs
decays $h \to \chi_1^0 \chi_1^0$ for $M_h = 90$ GeV and $\tb = 1.6$,
(maximal mixing conditions). The dash--dotted lines represent the
boundary of $[\mu, M_2]$ values after the LEP1.5 results are taken into
account. }
\end{figure}

\newpage
\begin{figure}[htbp]
\centerline{\psfig{figure=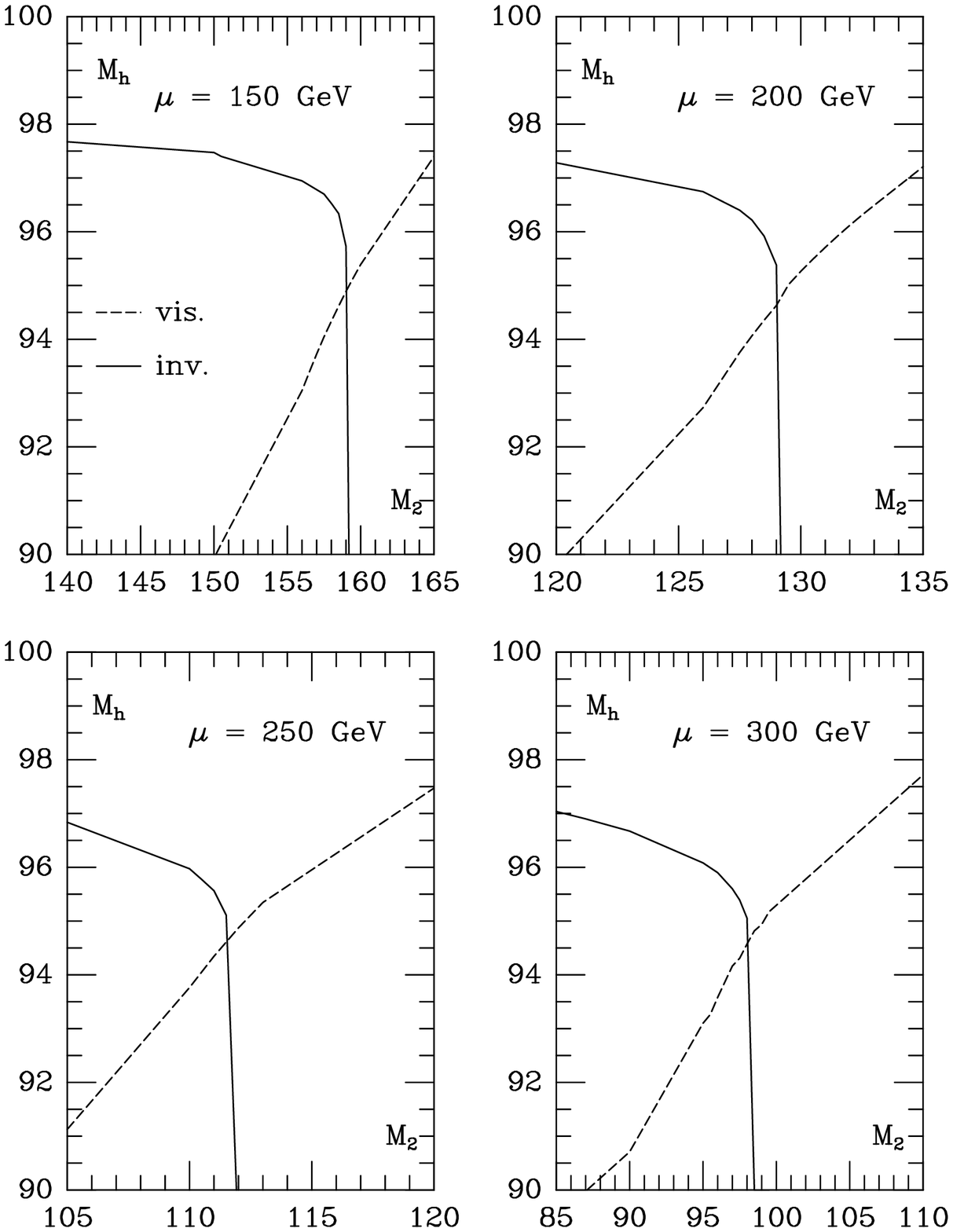,height=19cm,width=18cm}}
\vspace*{-1.2cm}
\noindent {\bf Fig.~3b:} {\it  Exclusion limits on the mass of the
lightest
CP--even Higgs boson $h$ for a set of $[\mu, M_2$] values
and $\tan \beta = 1.6$. The
full lines refer to the search based on the invisible $\chi_1^0
\chi_1^0$ decay channel, the dashed lines to $b\bar{b}$ decays of $h$.
The invisible decays are dominant whenever this channel is open, the
fast fall--off corresponds to the kinematical boundary.}
\end{figure}

\end{document}